\documentclass[twocolumn,showpacs,aps,pre]{revtex4}
\usepackage{amssymb}
\usepackage{graphicx}

\begin{document}

\title{Solitons in the Salerno model with competing nonlinearities}
\author{J. Gomez-Garde\~{n}es$^{1,2,4}$, B. A. Malomed$^{3}$,
L. M. Flor\'{\i}a$^{1,2}$, and A. R. Bishop$^{4}$\\
$^{1}$Departamento de F\'{\i}sica de la Materia Condensada and Instituto de Biocomputaci\'on y F\'{\i}sica de los Sistemas Complejos,
Universidad de Zaragoza, E-50009 Zaragoza, Spain\\
$^{2}$Instituto de Ciencia de Materiales de Arag\'on, C.S.I.C. - Universidad de Zaragoza, E-50009 Zaragoza, Spain\\
$^{3}$Department of Interdisciplinary Studies, School of
Electrical Engineering, Faculty of Engineering, Tel Aviv 69978,
Israel\\
$^{4}$Theoretical Division and Center for Nonlinear Studies, Los
Alamos National Laboratory, Los Alamos, New Mexico 87545, USA}

\begin{abstract}
We consider a lattice equation (Salerno model) combining onsite
self-focusing and intersite self-defocusing cubic terms, which may
describe a Bose-Einstein condensate of dipolar atoms trapped in a
strong periodic potential. In the continuum approximation, the
model gives rise to solitons in a finite band of frequencies, with
\textrm{sech}-like solitons near one edge, and an exact peakon
solution at the other. A similar family of solitons is found in
the discrete system, including a peakon; beyond the peakon, the
family continues in the form of cuspons. Stability of the lattice
solitons is explored through computation of eigenvalues for small
perturbations, and by direct simulations. A small part of the
family is unstable (in that case, the discrete solitons transform
into robust pulsonic excitations); both peakons and cuspons are
stable. The Vakhitov-Kolokolov criterion precisely explains the
stability of regular solitons and peakons, but does not apply to
cuspons. In-phase and out-of-phase bound states of solitons are
also constructed. They exchange their stability at a point where the
bound solitons are peakons. Mobile solitons, composed of a moving
core and background, exist up to a critical value of the strength
of the self-defocusing intersite nonlinearity. Colliding solitons
always merge into a single pulse.
\end{abstract}


\maketitle

\section{Introduction}

It is commonly known that the dynamics of nonlinear lattices are
drastically different in generic nonintegrable systems, a
paradigmatic example being the discrete nonlinear Schr\"{o}dinger
equation (DNLS; see reviews \cite{Panos}) and in exceptional
integrable models, a famous example of the latter being the
Ablowitz-Ladik (AL) equation \cite{AL}. Only in the latter case
are exact solutions for soliton families available. In
nonintegrable systems solitons are sought in a numerical form
\cite{Panos} or by means of a variational approximation
\cite{VA}. As the difference between the DNLS and AL equations is
in the type of the nonlinear terms -- onsite or intersite -- it
was quite natural to introduce a combined model that includes the
cubic terms of both types, and thus allows one to consider a
continuous transition between the AL and DNLS equations. The
combined equation, known as the Salerno model (SM) \cite{Salerno},
is \begin{equation} i\dot{\Phi}_{n}=-\left( \Phi _{n+1}+\Phi
_{n-1}\right) \left( 1+\mu \left\vert \Phi _{n}\right\vert
^{2}\right) -2\nu \left\vert \Phi _{n}\right\vert ^{2}\Phi _{n},
\label{Salerno}
\end{equation} where $\Phi _{n}$ is the complex field amplitude at 
the $n$-th site of the lattice, the overdot stands for the time 
derivative, and real coefficients $ \mu $ and $\nu $ account for 
the nonlinearities of the AL and DNLS types, respectively.

The SM was studied in a number of other works, see Refs.
\cite{LANL}-\cite {MovingSolitonsDNLS} and references therein. It is known
that it conserves the norm and Hamiltonian,
\begin{equation}
{\cal N}=\frac{1}{\mu }\sum_{n}{\ln }\left( \left\vert 1+\mu |\Phi
_{n}|^{2}\right\vert \right) ,  \label{eq:Norm}
\end{equation}
\begin{eqnarray}
{\cal H}=\sum_{n}\left[ -\left( \Phi _{n}\Phi _{n+1}^{\ast }+\Phi
_{n+1}\Phi _{n}^{\ast }\right) -2\frac{\nu }{\mu }|\Phi
_{n}|^{2}\right.
\nonumber
\\
\left. +2\frac{\nu }{\mu ^{2}} {\ln }\left( \left\vert 1+\mu
|\Phi _{n}|^{2}\right\vert \right) \right]\;.
\label{eq:SalernoEnergy}
\end{eqnarray}

In the above-mentioned works, it has been demonstrated that Eq.
(\ref {Salerno}) gives rise to static (and, sometimes, moving
\cite {LANLcollision}-\cite{MovingSolitonsDNLS}) solitons at all 
values of the DNLS parameter $\nu $ , and all \emph{positive} values 
of the AL coefficient, $\mu $. If $\nu $ is negative, one can make it
positive by means of the \textit{staggering transformation}, $\Phi
_{n}\rightarrow (-1)^{n}\Phi _{n}$, and then setting $\nu \equiv +1$,
by the rescaling $\Phi _{n}\rightarrow \Phi _{n}/\sqrt{\nu }$
(unless $\nu =0$). However, the sign of $\mu $ cannot be altered.
In particular, the AL model proper ($\nu =0$) with $\mu <0$ does
not give rise to solitons. The latter circumstance suggests 
considering soliton dynamics in the SM with $\mu <0$, i.e., with
\textit{competing nonlinearities}, which is the subject of the
present work. In this connection, it is necessary to stress that
expressions (\ref{eq:Norm}) and (\ref{eq:SalernoEnergy}) for the
SM's dynamical invariants remain valid if $1+\mu |\Phi _{n}|^{2}$
takes negative values at some sites, due to $\mu <0$.

While the SM was originally introduced in a rather abstract context, it has
recently found direct physical realization, as an asymptotic form of the
Gross-Pitaevskii equation describing a Bose-Einstein condensate of bosonic
atoms with magnetic momentum trapped in a deep optical lattice \cite{Liu}.
In that case, the onsite nonlinearity is generated, as usual, by collisions
between atoms, while the intersite nonlinear terms account for the
long-range dipole-dipole interactions. Note that the latter interaction may
be attractive ($\mu >0$) or repulsive ($\mu <0$), if the external magnetic
field polarizes the atomic momentum along the lattice or perpendicular to
it, respectively.

This report is organized as follows. In Section \ref{sec:CA} we develop 
a continuum approximation (CA), and investigate the corresponding solitons 
in an analytical form. It is found that, although they might exist in a
semi-infinite band of frequencies, they actually occupy a finite band, with
an exact peakon solution at the edge of the band. A family of discrete
solitons is constructed by means of numerical-continuation methods in
Section \ref{sec:Stationary}. They form a family of regular solitons, 
including a peakon, similar to what was found in the CA, but the lattice 
solitons extend beyond the peakon in the form of cuspons. In Section 
\ref{subsec:Stability}, the soliton stability is explored by means of 
standard Floquet analysis (computation of perturbation eigenvalues or 
Floquet multipliers) and in direct simulations, with the conclusion that 
only a small part of the family is unstable. 
Two-soliton bound states are reported in Section \ref{subsec:Bound}, where 
it is demonstrated that stability exchange between in-phase and out-of-phase
states occurs at a point where the bound solitons are peakons. Moving
solitons are considered in Section \ref{sec:Mobile}, where it is found that 
they exist up to a critical strength of the intersite self-defocusing 
nonlinearity, and collisions between them always lead to fusion into a 
single soliton. The paper is concluded in Section \ref{sec:Conclusion}.

\section{Continuum limit}
\label{sec:CA}

To introduce the CA in Eq. (\ref{Salerno}), we define $\Phi
(x,t)\equiv e^{2it}\Psi (x,t)$, and expand $\Psi _{n\pm
1}\approx \Psi \pm \Psi _{x}+(1/2)\Psi _{xx}$, where $\Psi $ is
now treated as a function of the continuous coordinate $x$, which
coincides with $n$ when it takes integer values. After that, the
continuum counterpart of Eq. (\ref{Salerno}) is derived,
\begin{equation}
i\Psi _{t}=-2\left( 1-|\mu |\right) \left\vert
\Psi \right\vert ^{2}\Psi -\left( 1-\left\vert \mu \right\vert
\left\vert \Psi \right\vert ^{2}\right) \Psi _{xx}~,  \label{Psi}
\end{equation}
where we have set $\nu =+1$ and $\mu <0$, as stated above. Equation
(\ref{Psi} ) conserves the norm and Hamiltonian, which are
straightforward counterparts of expressions (\ref{eq:Norm}) and
(\ref{eq:SalernoEnergy}),
\begin{eqnarray}
{\cal N}_{\mathrm{cont}}=\frac{1}{\mu }\int_{-\infty
}^{+\infty }dx~{\ln }\left( \left\vert 1-|\mu ||\Psi
|^{2}\right\vert \right) ,  
\label{Ncont}
\\
{\cal H}_{\mathrm{cont}}=\int_{-\infty }^{+\infty }\left[\left\vert \Psi
_{x}\right\vert ^{2}+2\left( \frac{1}{|\mu |}-1\right) |\Psi
|^{2}\right.
\nonumber
\\
\left. +\frac{2}{ \mu ^{2}}{\ln }\left( \left\vert 1-|\mu ||\Psi
|^{2}\right\vert \right) \right] .
\label{Hcont}
\end{eqnarray}
Soliton solutions to Eq. (\ref{Psi}) are sought as $\Psi
=e^{-i\omega t}U(x)$, with a real function $U$ obeying the
equation
\begin{equation}
\frac{d^{2}U}{dx^{2}}=-\frac{\omega
+2\left( 1-|\mu |\right) U^{2}}{1-|\mu |U^{2}}U,  \label{U}
\end{equation}
which may give rise to solitons, provided that $\omega <0$ and $|\mu |<1$.
(The absence of solitons for $|\mu |>1$ implies that if the
intersite self-defocusing, accounted for by $\mu <0$, is stronger
than the onsite self-focusing, the self-trapping of solitons is
impossible in the CA). Equation (\ref{U}) can be cast in the form
$\left( (1/|\mu |)-1\right) ^{-1}U_{xx}^{\prime \prime
}=-W^{\prime }(U)$, where the effective potential is
\begin{equation}
W=-\frac{1}{2}U^{2}-\frac{1-\Omega }{2|\mu |}\ln
\left( 1-|\mu |U^{2}\right) ,\;\;\Omega \equiv \frac{\mu \omega
}{1-|\mu |};  \label{potential}
\end{equation}
the expansion of the potential (\ref{potential}) for $U^{2}\rightarrow 0$ is
$W\approx \left[ -\Omega U^{2}+|\mu |\left( 1-\Omega \right)
U^{4}\right] /2$ . This form of the equation shows that solitons
exist in a \emph{finite } band of frequencies, $0<\Omega <1$,
rather than in the entire semi-infinite band, $\Omega >0$, where
the linearization of Eq. (\ref{U}) produces exponentially decaying
solutions that could serve as the solitons' tails. The reduction of
the semi-infinite band to a finite one is typical for soliton
families in models with competing nonlinearities, such as the
cubic-quintic NLS equation \cite{Pushkarov}. Further, it follows
from the divergence of potential (\ref{potential}) at
$U^{2}=1/|\mu |$ that the solitons's amplitude $A$, which is a
monotonously increasing function of $ \Omega $, is smaller than
$1/\sqrt{|\mu |}$ for $0<\Omega <1$, and $A=1/\sqrt{ |\mu |}$ at
$\Omega =1$.

Solitons can be found in an explicit form near the edges of the
existence band: at small $|\omega |$ (i.e., small $\Omega $),
$U(x)\approx \sqrt{ |\omega |/\left( 1-|\mu |\right)
}\mathrm{sech}\left( \sqrt{2|\omega |} x\right) $, while precisely
at the opposite edge of the band, $\omega =1-1/|\mu |$ (i.e.,
$\Omega =1$), the exact solution is a \textit{peakon},
\begin{equation}
U_{\mathrm{peakon}}=\left( 1/\sqrt{|\mu |}\right)
\exp \left( -\sqrt{\left( 1/|\mu |\right) -1}|x|\right) .
\label{peakon}
\end{equation}
In other words, at a given frequency $\omega $, the peakon solution is 
found at
\begin{equation}
|\mu |=\left\vert \mu _{p}\right\vert \equiv
1/\left( 1-\omega \right) .
\label{mu_p}
\end{equation}
Note that norm (\ref{Ncont}) of the peakon is $\pi ^{2}/[6\sqrt{|\mu
|(1-|\mu |)}]$, and its energy is also finite. Close to this point,
i.e., for $0<1-\Omega \ll 1$, the solution is different from the
limiting form (\ref {peakon}) in a narrow interval $|x|\lesssim
\sqrt{|\mu |/\left( 1-|\mu |\right) }(1-\Omega )$, where the peak
is smoothed.

Finally, the CA based on Eq. (\ref{Psi}) is valid if the intrinsic
scale of all continuum solutions, that may be estimated through
the curvature of the soliton's profile at $x=0$ as $l\sim
1/\sqrt{\left\vert U_{xx}^{\prime \prime }/U\right\vert }$, is
large, $l\gg 1$ (recall the lattice spacing is $ 1$ in the present
notation). According to Eq. (\ref{peakon}), the latter condition
implies $(1/|\mu |)-1\ll 1$ (i.e., strictly speaking, the CA
applies in the case when the competing nonlinearities in the SM
nearly cancel each other).

It is relevant to note that, in the usually considered
``non-competitive" version of the SM, with $\mu
>0$ (i.e., self-focusing intersite nonlinearity), the CA yields
solitons in the entire semi-infinite band, $\omega <0$.

\section{Stationary discrete solitons}
\label{sec:Stationary}

In order to find discrete solitons in a numerical form, we looked for 
solutions to Eq. (\ref{Salerno}) which are localized and time periodic 
with frequency $\omega _{b}=2\pi/T_{b}$ \cite{Footnote} 
(that is related to $\omega $ in the continuum equation by $\omega
_{b}\equiv \omega -2$). Soliton solutions of this form are widely 
known for the DNLS limit ($\mu=0$) and hence it is posible to make a 
numerical continuation of such solutions for $\mu<0$ by adiabatic 
changes of the model parameter $\mu$ and successive applications of the 
shooting method \cite{JLMarin,Cretegny}. For this purpose we consider soliton 
solutions of the above form as fixed points of the map 
${\cal T}_{T_{b},\mu}$ defined as
\begin{equation}
{\cal T}_{T_{b},\mu}\Phi_{n}(t)=\Phi_{n}(t+T_{b})\;,
\end{equation}
{\em i.e} ${\cal T}_{T_{b,\mu}}$ is the time-evolution operator 
over $T_{b}$ of dynamics dictated by Eq. (\ref{Salerno}). 
Then, using usual techniques for finding fixed points of maps 
(such as {\em e.g.} the Newton-Raphson algorithm) we can find the 
numerically exact soliton for a given frequency $\omega_{b}$ and $\mu$.  
In general all the soliton solutions were computed starting from the 
DNLS limit, $\mu =0$, and increasing $|\mu |$ at a fixed value 
of $\omega _{b}$. The continuations were performed using the 
shooting method, with an increment $\delta (|\mu |)=10^{-2}$ 
at each step, or smaller if higher accuracy was needed.

As shown in the previous section, the soliton family in the
continuum equation (\ref{Psi}) ends with the peakon\ solution
(\ref{peakon}). To compare the numerically determined shape of the
discrete solitons with the feasible peakon limit, we fitted the
solitons' tails to the asymptotic form, $|\Phi _{n}|=A\exp
(-\Gamma \left( \left\vert n-n_{0}\right\vert \right) )$, with
constant $A$, $\Gamma $, and $n_{0}$, which follows from the
linearized equation (\ref{Salerno}) for large $|n|$. This
procedure yielded the decay rate, $\Gamma =\Gamma (\mu ,\omega
_{b})$, amplitude, $A=A(\mu ,\omega _{b})$ (and the center's
position $n_{0}$), as functions of parameters $\mu $ and $\omega _{b}$ 
of the soliton family. Once $A(\mu ,\omega _{b})$ and $n_{0}$ 
were found, we defined $\gamma (\mu ,\omega _{b})$ 
$\equiv A-|\Phi _{n_{0}}|$ to measure a deviation of the true 
discrete soliton from a conjectured peakon shape obtained by formal 
extension of the tail inward.

\begin{figure}[tbp]
\begin{center}
\includegraphics[angle=-0,width=.48\textwidth]{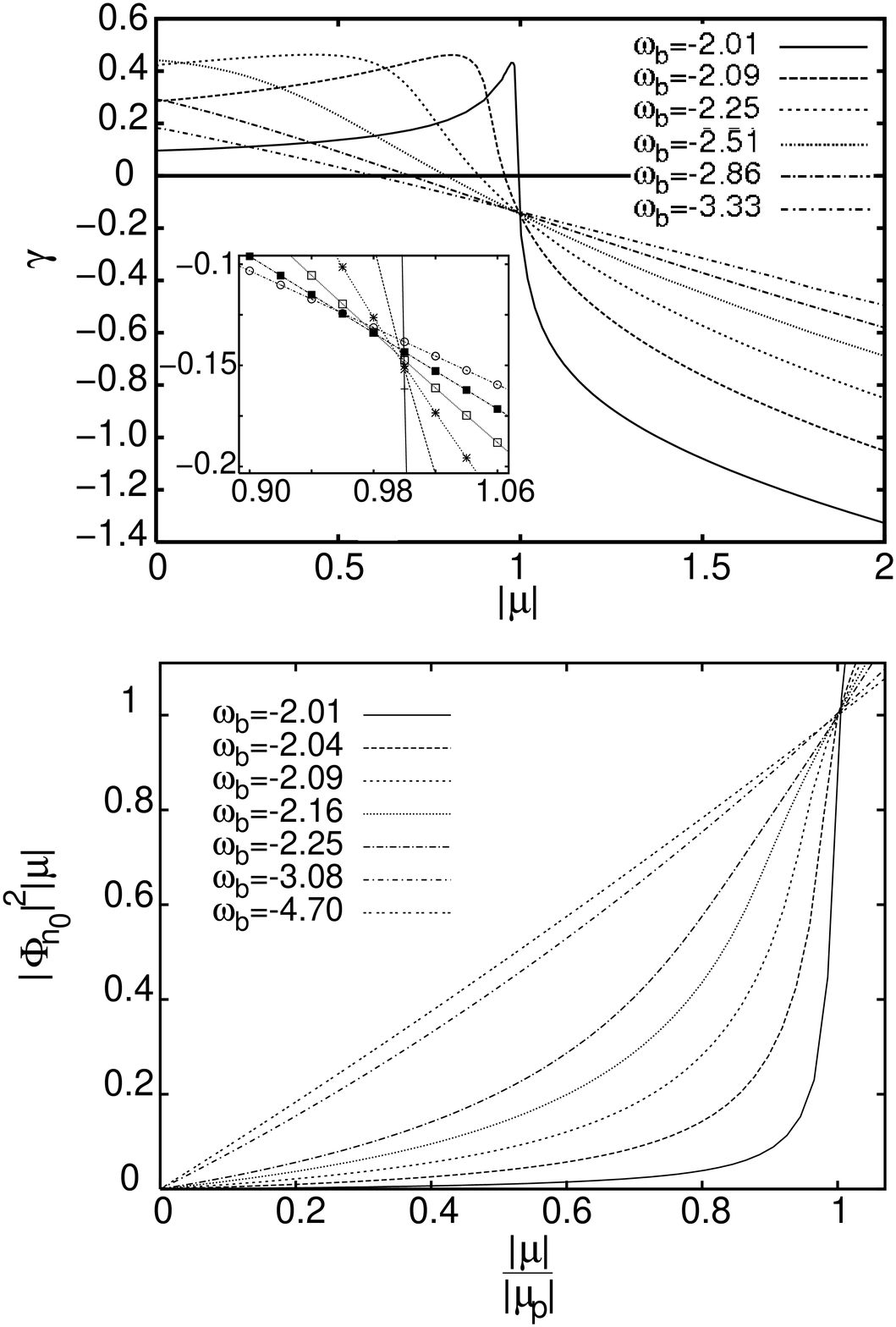} 
\end{center}
\caption{ \textbf{(}a\textbf{)} The mismatch with the peakon shape,
$\protect\gamma $ , as a function of $|\protect\mu |$, for
discrete solitons found at different frequencies $\protect\omega
_{b}$. (Note in the inset that there is no common intersection of 
all the curves). \textbf{(}b\textbf{)} The soliton's amplitude vs.
$|\protect\mu |$. The axes are rescaled to show that the amplitude
of the peakon solutions (attained at $|\protect\mu |=|\protect \mu
_{p}|$) are equal to $1/\protect\sqrt{|\protect\mu |}$, as
predicted by the continuum approximation.}
\label{fig:1}
\end{figure}

In Fig. \ref{fig:1}(a) we show the evolution of $\gamma $ produced
by several continuations of the lattice-soliton solutions (at
different frequencies $\omega _{b}$). We define $\mu _{p}(\omega
_{b})$ as a value of $ \mu $ at which an exact discrete peakon of 
internal frequency $\omega_{b}$ is found, that we realize as 
vanishing of $\gamma \left( \mu ,\omega_{b}\right) $ at $\mu =\mu _{p}$. 
In Fig. \ref{fig:1}(b) we plot the evolution of the soliton's amplitude 
as the continuation is performed. It is observed that the amplitude
increases with $|\mu |$, reaching the predicted value,
$1/\sqrt{|\mu |}$, at the exact peakon solution.

A noteworthy result, evident from Fig. \ref{fig:1}, is the persistence of
the discrete solitons \emph{beyond the peakon limit} (which means
continuability of the solutions to $\gamma <0$). The apparent intersection
of different curves at one point in Fig\ref{fig:1}(a) is a 
spurious feature (see the inset in the figure): an accurate consideration 
shows that the curves actually intersect at close but different points. 
In contrast, the intersection of the curves in Fig. \ref{fig:1}(b) indeed 
happens at a single point, which corresponds to the discrete solitons taking 
the peakon shape.

\begin{figure}[tbp]
\begin{center}
\begin{tabular}{cc}
\includegraphics[angle=-0,width=.45\textwidth]{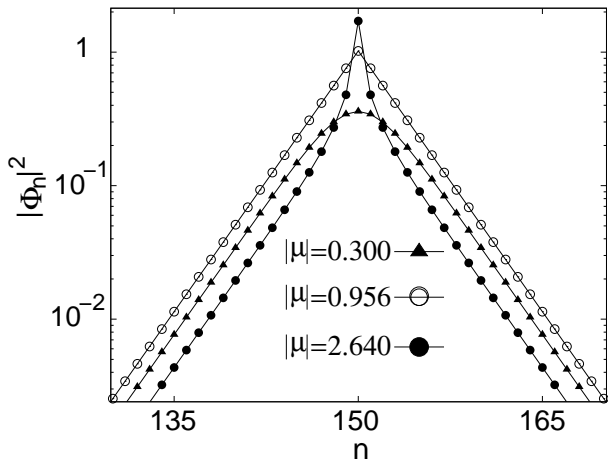} 
\end{tabular} 
\end{center}
\caption{Generic examples of three different types of discrete 
solitons, for $\protect\omega_{b}=-2.091$: a quasi-continuous 
$ \mathrm{sech}$-like solution at $|\protect\mu |=0.3$, 
a peakon at $|\protect \mu |=0.956$, and a cuspon at 
$|\protect\mu |=2.64$.}
\label{fig:2}
\end{figure}

Figure \ref{fig:2} displays typical examples of the numerically
found discrete solitons. It demonstrates that the solutions
corresponding to $ \gamma <0$ are \textit{cuspons}, with a
super-exponential shape, that do not exist in the continuum
equation (\ref{Psi}). The discrete character of the SM with the
competing nonlinearities allows this new type of solution (as
happens with the quasi-collapsing states in the standard DNLS
equation in two dimensions \cite{quasicollapse}). Cuspon solutions
continue into the region of $|\mu |>1$, where the CA yields no
solitons, but, due to the sharp change of the solution with the
increase of $|\mu |$, finding numerical solutions at larger values
of $|\mu |$ becomes increasingly more difficult.

\begin{figure*}[tbp]
\begin{center}
\begin{tabular}{c}
\includegraphics[angle=-0,width=.95\textwidth]{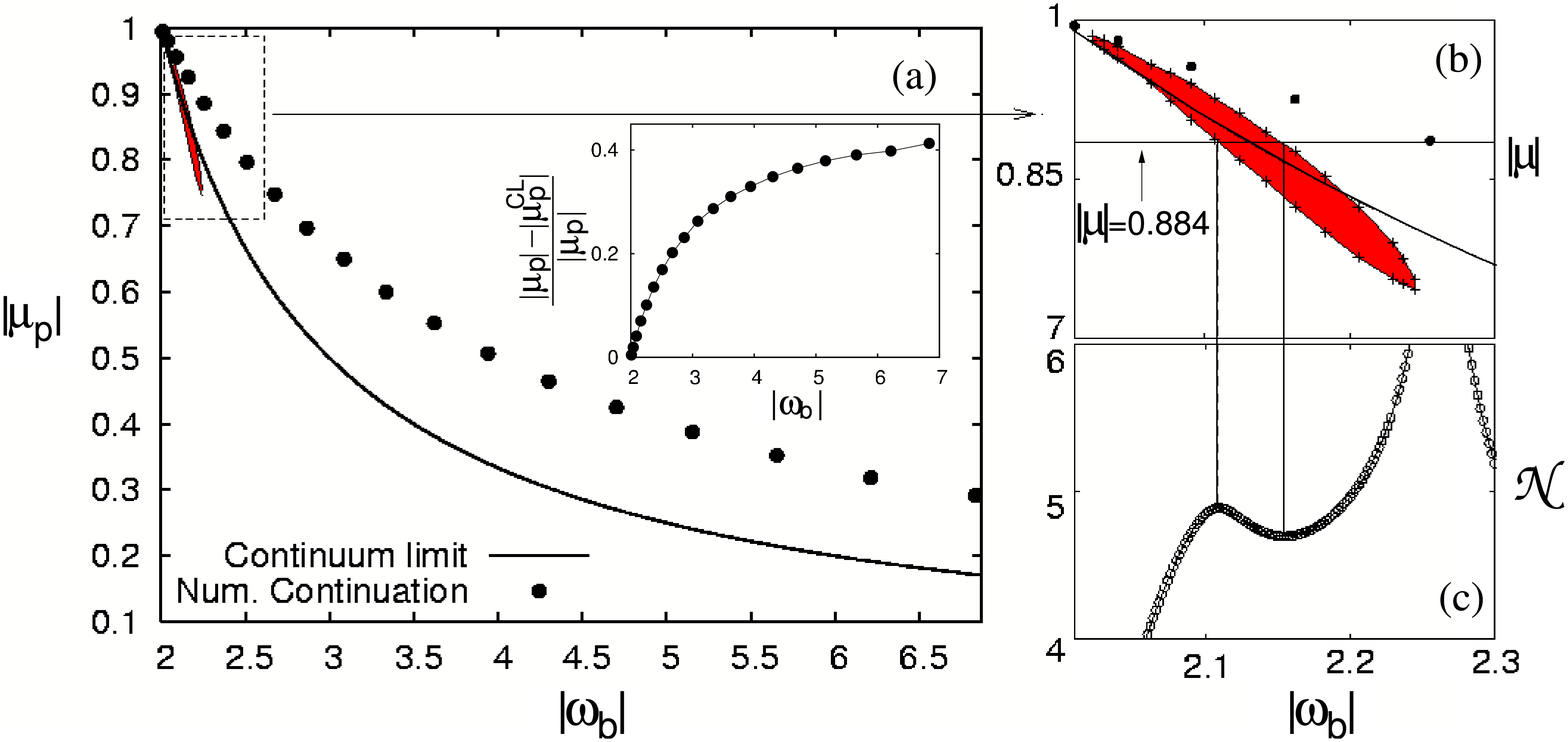}
\end{tabular}
\end{center}
\caption{(Color online). \textbf{(}a\textbf{)} The value of $\left\vert
\protect\mu  _{p}\right\vert $, at which the soliton assumes the
peakon shape: the prediction of the continuum approximation, Eq.
(\protect\ref{mu_p}) (solid curve), and numerical results for
discrete solitons (dots). Also shown is a small region where the
discrete solitons are found to be unstable (for that purpose, the
vertical axis shows $|\protect\mu |$, rather than $|\protect\mu
_{p}|$). The inset displays the relative difference between the
numerically found values of $|\protect\mu _{p}|$ and the
prediction, $|\protect\mu  _{p}^{CL}|$, provided by the continuum
approximation. \textbf{(}b\textbf{)} Zoom of the area in the
$\left( |\omega _{b}|,\left\vert \protect\mu
_{p}\right\vert \right) $ plane where the instability island is
located. \textbf{(}c\textbf{)} Norm of the discrete solitons vs.
the frequency, for $| \protect\mu |=0.884$.}
\label{fig:3}
\end{figure*}

In Fig. \ref{fig:3}(a) we compare the line of the existence of the peakons
in the continuum limit, and the actual location of discrete peakons. It is
seen that the agreement between the CA and numerical findings is good for
smaller $\left\vert \omega _{b}\right\vert $ (in this case, the discrete
solitons are broad), while at larger $\left\vert \omega _{b}\right\vert $
the discrete solitons are narrow, hence the agreement with the CA
deteriorates.

\subsection{Stability of the solitons}
\label{subsec:Stability}

\begin{figure}[tbp]
\begin{center}
\begin{tabular}{cc}
\includegraphics[angle=-0,width=.45\textwidth]{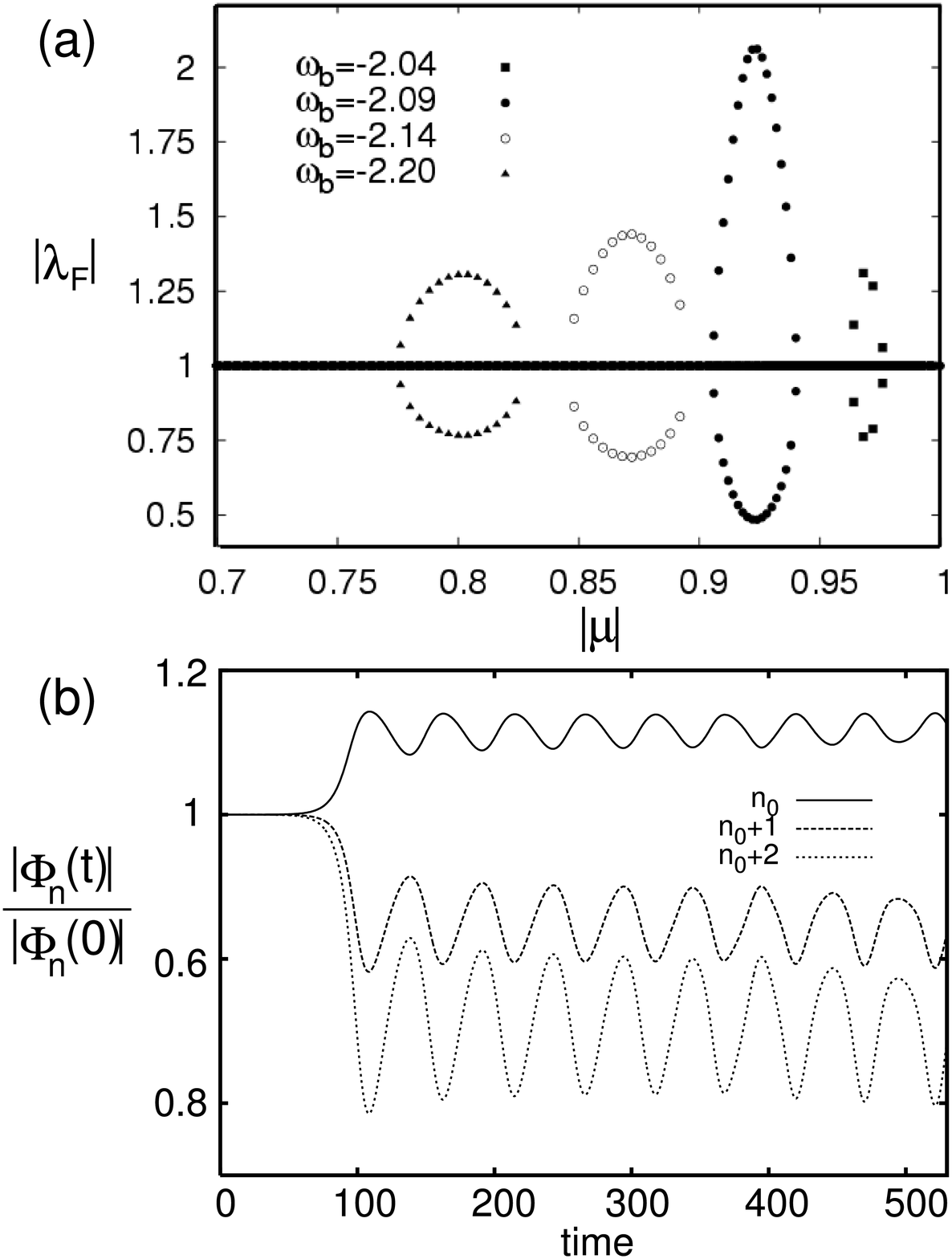} 
\end{tabular}
\end{center}
\caption{\textbf{(}a\textbf{)} The
absolute value of the Floquet multiplier, $\left\vert
\protect\lambda _{F}\right\vert $, for the linearization of
perturbations around discrete solitons, is shown vs. $|\protect\mu
|$, at several fixed values of the frequency (which are chosen so
as to make the instability intervals well separated). The soliton
is unstable if $ \left\vert \protect\lambda
_{F}\right\vert \neq 1$. \textbf{(}b\textbf{)} A robust pulson generated
from an unstable soliton at $|\protect\mu |=0.922$.}
\label{fig:4}
\end{figure}

We have performed stability analysis of the discrete solitons by
computing eigenvalues (Floquet multipliers, $\lambda _{F}$) for
modes of small perturbations \cite{Floria}, within the framework of the
linearized equation of the SM (\ref{Salerno}).  The integration of 
a basis of initial perturbations over a period $T_{b}$ of the discrete 
soliton gives the Floquet matrix that in our case is real and symplectic 
so that its eigenvalues come in quadruplets $\{\lambda _{F},\; 
1/\lambda _{F},\; \overline{\lambda}_{F},\;1/\overline{\lambda}_{F}\}$. 
Then, the discrete soliton solution to Eq.(\ref{Salerno}) is linearly 
stable if all the Floquet multipliers lie on the unit circle of the complex 
plane, $|\lambda _{F}|=1$. It is found that the solitons are linearly 
stable along the whole continuation, except for a relatively small region, 
as shown in Fig. \ref{fig:4}(a). 
The entire instability island in the $(|\omega _{b}|,|\mu |)$ 
plane is displayed in Fig. \ref{fig:3}(b). Note, in particular, 
that the peakon and cuspon solutions are stable. The stability 
of the discrete solitons was also checked by direct simulations 
of perturbed solitons, using the full equation (\ref{Salerno}), 
which corroborated the predictions of the linear analysis.

Direct simulations of the evolution of perturbed unstable solitons, a
typical example of which is displayed in Fig. \ref{fig:4}(b), show that,
after a transient stage, a localized pulson (showing simultaneous width 
and amplitude oscillations) is formed. The pulsons are (quasi-)periodic 
in time, and persist indefinitely. This behavior resembles that found 
in the ordinary two-dimensional DNLS equation in quasi-collapsing states 
\cite{quasicollapse}.

A necessary stability condition for soliton families in models of the NLS
type may be provided by the Vakhitov-Kolokolov (VK) criterion \cite{VK}: if
the norm ${\cal N}$ of the soliton is known as a function of its frequency 
$\omega_{b}$, the solitons can be stable against small perturbations with real
eigenvalues, provided that $d{\cal N}/{d}${$\omega _{b}<0$}. Although the
applicability of the VK criterion to the present model has not been proven
(and counter-examples are known, when solitons predicted by the criterion 
to be unstable are actually stable \cite{KronigPenney}), it is relevant to 
test the criterion here, numerically computing 
${\cal N}\left( \omega _{b}\right)$ according to Eq. (\ref{eq:Norm}). 
The result is that the VK\ criterion \emph{precisely} explains the stability 
and instability of the discrete solitons, except for the cuspons (see below), 
as shown in Fig. \ref{fig:3}(c).

\begin{figure}[tbp]
\begin{center}
\begin{tabular}{c}
\includegraphics[angle=-0,width=.45\textwidth]{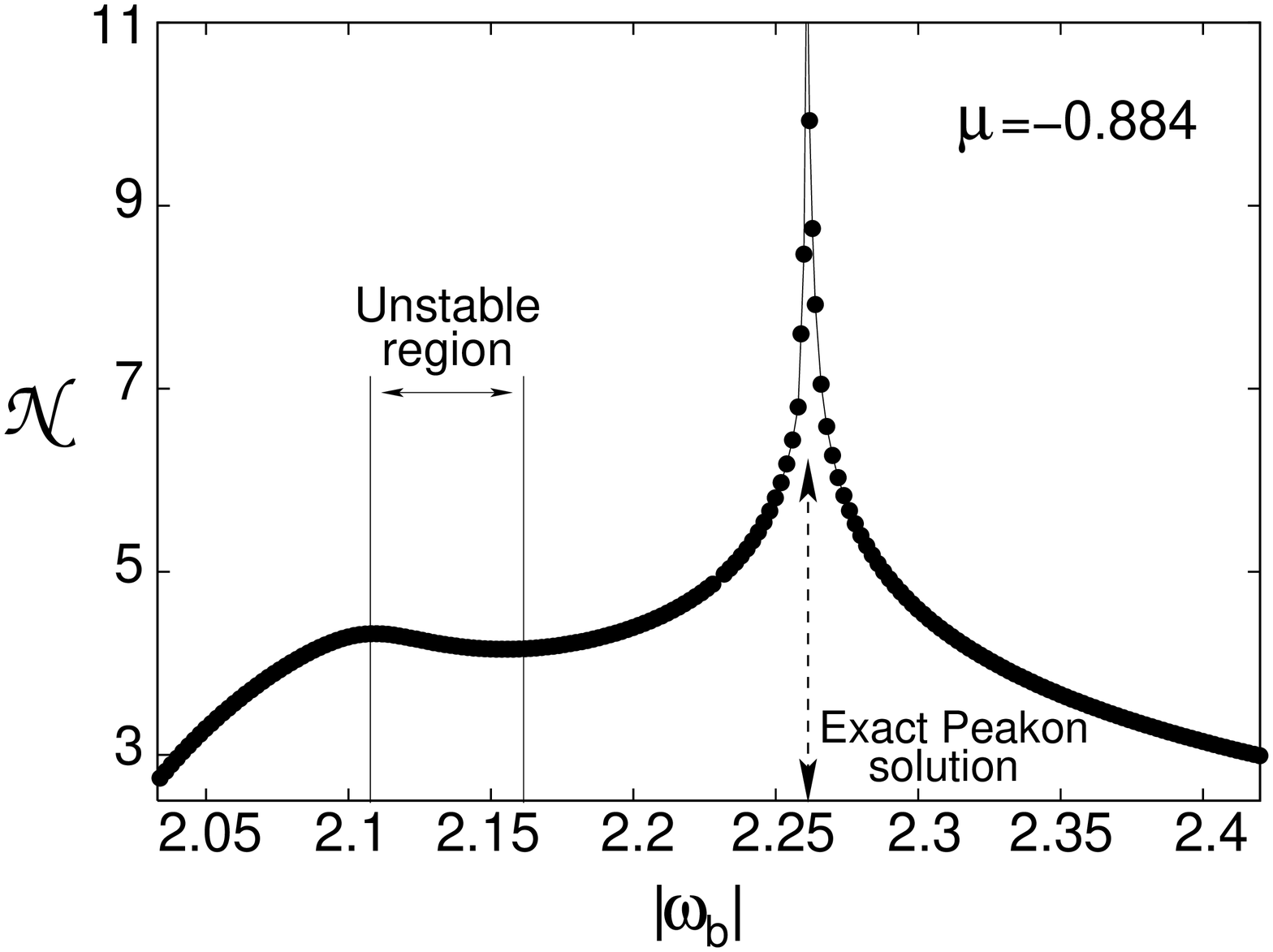}
\end{tabular}
\end{center}
\caption{The norm of the discrete solitons vs. the frequency, for
$|\protect \mu |=0.884$.}
\label{fig:5}
\end{figure}

A noteworthy feature of the ${\cal N}\left( \omega _{b}\right) $
dependence is a divergence of the total norm due to the infinite
contribution of the central site to expression (\ref{eq:Norm}) in
the case of the exact peakon solution, with $\left\vert \Phi
_{n_{0}}\right\vert ^{2}=1/|\mu |$. An example of the $ {\cal N}\left(
\omega _{b}\right) $ dependence showing the divergence is plotted
in Fig. \ref{fig:5}. As concerns the cuspons, whose amplitude
exceeds the critical value, $1/\sqrt{|\mu |}$, the norm
(\ref{eq:Norm}) converges for them, and features a positive slope
(see Fig. \ref{fig:5}), $d{\cal N}/d\omega _{b}>0$. The VK criterion
predicts instability in this case; however, the direct computation
of the Floquet multipliers (see above), as well as direct
simulations, reveal \emph{no instability} of the cuspons. Thus,
while the VK criterion is perfectly correct for regular solitons
and peakons in the present model, it is irrelevant for cuspons, cf. 
the situation in Ref. \cite{KronigPenney}.

\subsection{Bound states of solitons and their stability}
\label{subsec:Bound}

We have explored bound states of discrete solitons in Eq.
(\ref{Salerno}). For this purpose, we performed numerical
continuation in $\mu $, starting with the well known bound states
of the standard DNLS equation, at $\mu =0$. In that limit, two
different types of bound states are known, in-phase and $ \pi
$-out-of-phase ones, which are represented, respectively, by even
and odd solutions. It is well known that only the states of
the latter type are stable \cite{Todd}.

\begin{figure*}[tbp]
\begin{center}
\begin{tabular}{cc}
(a) \includegraphics[angle=-0,width=.44\textwidth]{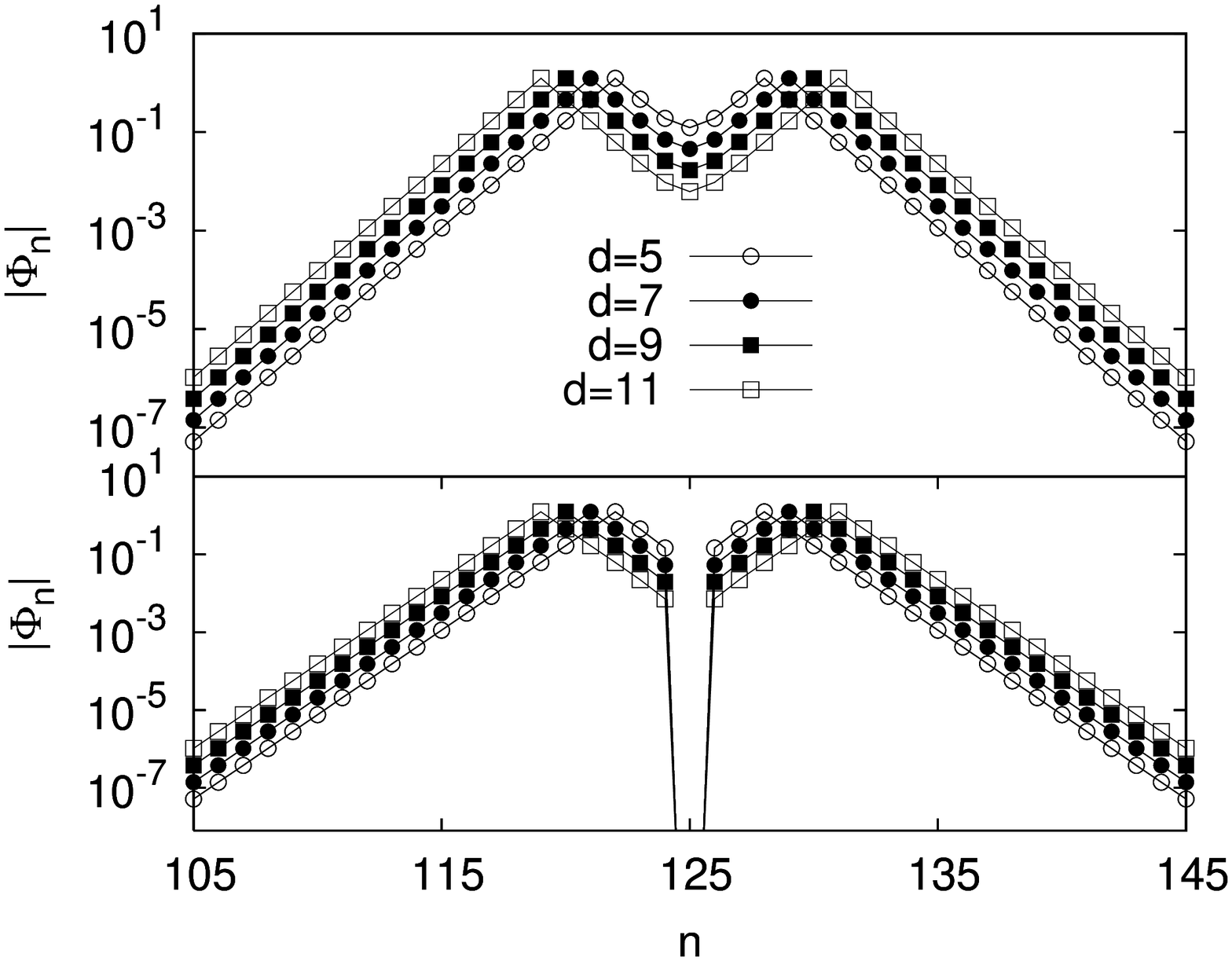} &
\includegraphics[angle=-0,width=.44\textwidth]{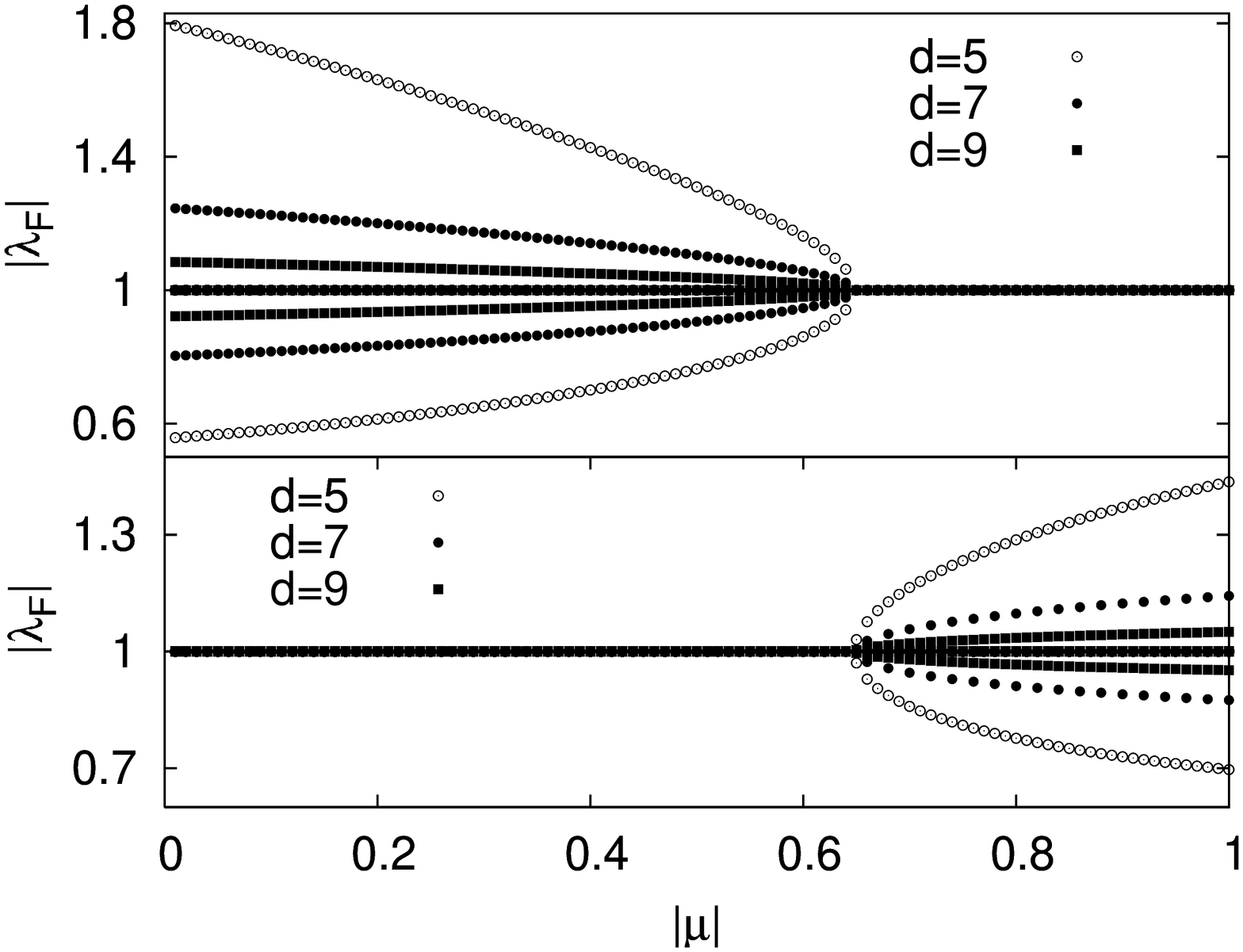} (b)
\end{tabular}
\end{center}
\caption{(a) Profiles of typical
in-phase (top) and out-of-phase (bottom) bound states of two
peakons, with different distances between their centers, at
$\protect\omega _{b}=-3.086$ and $|\protect\mu |=|\protect\mu
_{p}|=0.645$ . (b) Absolute values of the Floquet multiplies that
determine the stability of three bound states, with the same fixed
frequency, $\protect\omega  _{b}=-3.086$, and different
separations between the solitons. The in-phase (top) and
out-of-phase (bottom) bound states are stabilized and
destabilized, respectively, at the point where the bound solitons
are peakons, see panel (a). Unstable states are less unstable
(with smaller absolute values of the Floquet multipliers
accounting for the instability) if the distance between the
solitons is larger.}
\label{fig:7}
\end{figure*}

The numerical continuation of soliton bound states was performed for 
pairs of identical discrete solitons of a given frequency $\omega_{b}$ 
and different distances between them. The continuation of in- and 
out-of-phase bound states gives bound states of peakons, see Fig. 
\ref{fig:7}(a). The latter solution is found at exactly the same
value, $\mu =\mu _{p}(\omega _{b})$, which gives rise to the 
single peakon. We have also examined the stability eigenvalues 
for the computed solutions. A remarkable feature of the bound states 
observed with increase of $|\mu|$ is the \emph{stability interchange} 
between the in-phase and out-of-phase states, as shown in Fig. 
\ref{fig:7}(b), which occurs precisely at $\mu =\mu_{p}(\omega _{b})$, 
regardless of the separation between the bound solitons.

\section{Moving solitons}
\label{sec:Mobile}

The integrable AL model gives rise to both static and moving solitons. 
In the DNLS equation, moving solitons \cite{Mussl,Papa} and
collisions between them \cite{Papa} were also studied, using
various methods. Soliton motion was also a subject of analysis in
the SM, which was usually treated, in this context, as a perturbed
version of the AL equation \cite {LANLcollision,Dmitriev}.

\begin{figure}[tbp]
\begin{center}
\begin{tabular}{c}
\includegraphics[angle=-90,width=.45\textwidth]{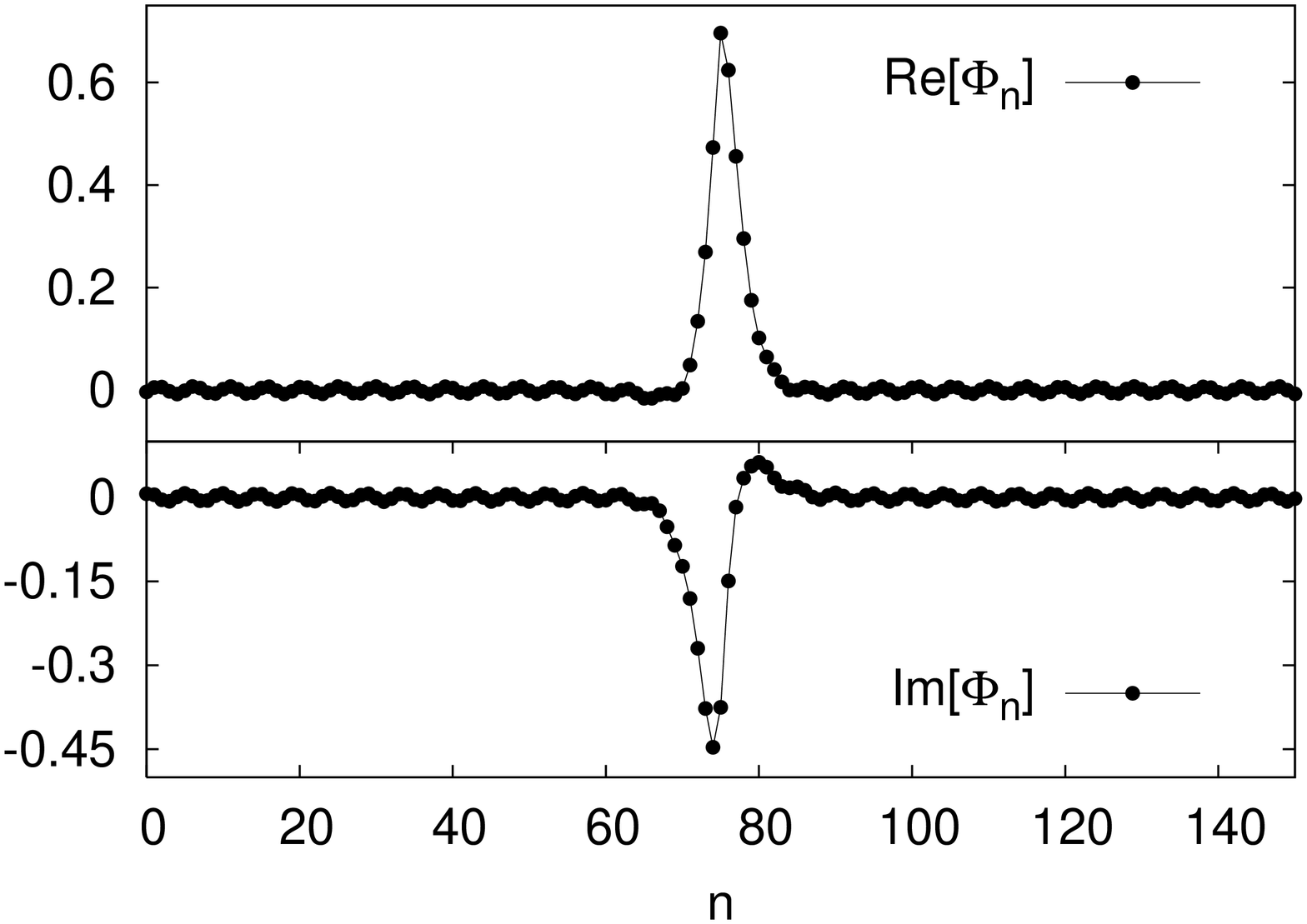}
\end{tabular}
\end{center}
\caption{The real and imaginary parts of the lattice wave field in a moving
discrete soliton, for $\protect\omega _{b}=-2.24$ and $\protect\mu =-0.7$.}
\label{fig:6pre}
\end{figure}

As a part of the present work, we have performed numerical
continuation of mobile discrete solitons from the DNLS limit,
where, in turn, they were earlier obtained by means of a
continuation procedure initiated with the AL solitons
\cite{MovingSolitonsDNLS}. The continuation procedure is the 
natural extension of the one applied to static discrete 
solitons ({\em i.e.} based on computing fixed points of maps 
by means of the shooting method) and allows obtaining mobile 
solutions whose profiles are repeated (translated a number of sites) 
after an integer number of periods of the internal oscillation. 
These states are composed of a traveling localized core and 
an extended background, 
$\Phi _{n}=\Phi_{n}^{\mathrm{core} }+\Phi _{n}^{\mathrm{bckg}}$, 
see Fig. \ref{fig:6pre}. The background is a superposition of 
nonlinear plane waves, and its amplitude is related to the height 
of the corresponding Peierls-Nabarro (PN) barrier, which is 
defined as the energy difference between two static solutions for the
soliton, with a fixed frequency $\omega _{b}$, one centered at a
lattice site, with $ n_{0}=n$, and the other at an intersite
position, with $n_{0}=n\pm 1/2$.

\begin{figure}[tbp]
\begin{center}
\begin{tabular}{c}
\includegraphics[angle=-0,width=.49\textwidth]{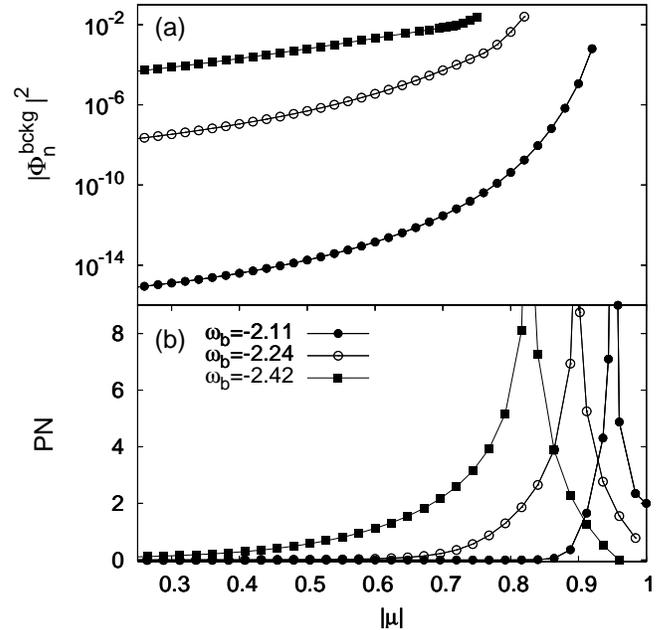}
\end{tabular}
\end{center}
\caption{The background amplitude (a) and the height of the Peierls-Nabarro
barrier (b), as functions of $|\protect\mu |$, for three mobile discrete
solitons.}
\label{fig:6}
\end{figure}

The result obtained along these lines in the present model is that
the mobile solitons can only be continued up to a certain critical
value, $\mu =$ $\mu _{c}(\omega _{b})$, close to but smaller in
absolute value than $ \mu _{p}(\omega _{b})$, at which the
static discrete soliton becomes a peakon. 
The Floquet stability analysis reveals that the extended background of the 
mobile solitons is subjected to modulational instability. (However, this is 
too weak to manifest itself in the simulations and it is only noticeable 
by looking at the Floquet spectra when the amplitude of the background is 
very high). On the the other hand we do not observe any localized eigenvector 
with eigenvalue $|\lambda_{F}|>1$ and thus the core is not affected by any 
unstable perturbation. The stability of mobile solutions is corroborated 
when simulations of the dynamics are performed allowing for interesting 
numerical experiments (see below). The background amplitude is a growing 
function of $|\mu |$ having a very sharp increase when $|\mu |$ approaches 
$\mu =$ $\mu _{c}(\omega _{b})$, see Fig.\ref{fig:6}(a). This behavior of 
the background amplitude suggests that the PN barrier also grows with 
$|\mu |$ and becomes very high near the critical point. 
To check this expectation, we have computed the height of the PN barrier 
for the same frequencies $\omega _{b}$ for which the mobile solitons 
were numerically calculated, using the energy definition as in Eq. 
(\ref{eq:SalernoEnergy}). Figure \ref{fig:6}(b) confirms that the PN 
barrier dramatically increases when the continuation approaches the 
critical point, $\mu =$ $\mu _{c}(\omega _{b})$, although the PN barrier 
diverges not exactly at this point, but rather at $\mu =\mu _{p}(\omega
_{b}) $, where the soliton assumes the peakon shape.

\begin{figure}[tbp]
\begin{center}
\begin{tabular}{c}
\includegraphics[angle=-0,width=.40\textwidth]{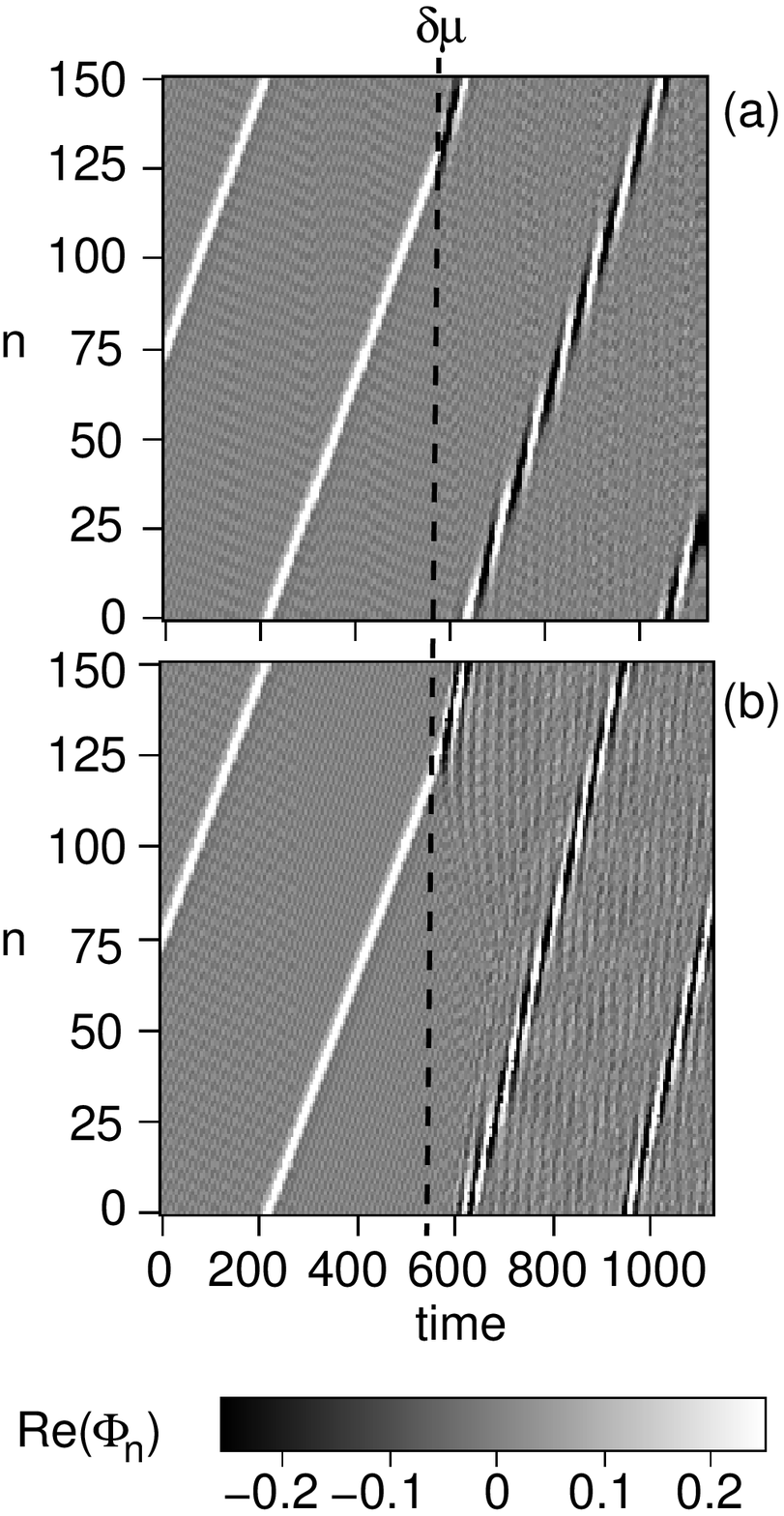}
\end{tabular}
\end{center}
\caption{The time evolution of the real part of the lattice wave
field in a mobile state. Initially, a moving soliton is created
corresponding to $| \protect\mu |=0.8$ and $\protect\omega
_{b}=-2.24$. Then, $\protect\mu $ jumps instantaneously to
$\protect\mu ^{\prime }=\protect\mu +\protect\delta \protect\mu $.
After the jump, followed by emission of some transient radiation,
the moving core becomes broader and starts to move faster. In (a)
$|\protect\mu ^{\prime }|=0.84$, and in (b) $|\protect\mu ^{\prime
}|=0.89$.}
\label{fig:6bis}
\end{figure}

The strong dependence of the PN barrier on $\mu $ suggests a numerical
experiment to test the behavior of mobile solitons when the lattice's
pinning force suddenly changes. To this end, we took an initial mobile
soliton at values of $\mu $ and $\omega _{b}$ for which the PN barrier is
low. Then we monitored the evolution of the moving soliton following an
instantaneous change in the nonlinearity, $\mu \rightarrow \mu +\delta \mu \equiv
\mu ^{\prime }$, which makes the PN barrier essentially higher than
experienced by the original soliton. The numerical experiments are
illustrated by Fig. \ref{fig:6bis}. We observe that the core of the mobile
lattice soliton does not become pinned due to the increase of the PN 
barrier, but rather accommodates itself, with some radiation loss, 
into a broader state with a smaller amplitude, so that the PN barrier, 
as experienced by the new state for $\mu =\mu ^{^{\prime }}$, is low 
enough to allow the soliton to remain mobile. Besides that, we observe an 
increment in the core's velocity, so that the larger the jump of the 
PN-barrier's height the faster is the new moving state.
The fact that the sudden increase of the PN barrier does not prevent 
the motion of the soliton reveals, on one hand, that the relation between 
PN barrier and mobility is far from trivial, and on the other hand, that 
mobility is quite a robust feature.

Finally, we simulated collisions between identical lattice
solitons moving in opposite directions. The results show that the
colliding solitons \emph{ always} merge into a single localized
state, which subsequently features intrinsic pulsations. If the PN
barrier is low, the emerging pulse can itself move in a chaotic
way, due to interaction with the lattice phonon field (radiation)
generated in the course of the collision. On the contrary, for
values of $\mu $ and $\omega _{b}$ at which the original solitons
experience a high PN barrier, the finally generated single soliton
is always strongly pinned.

\begin{figure}[tbp]
\begin{center}
\begin{tabular}{cc}
\includegraphics[angle=-0,width=.45\textwidth]{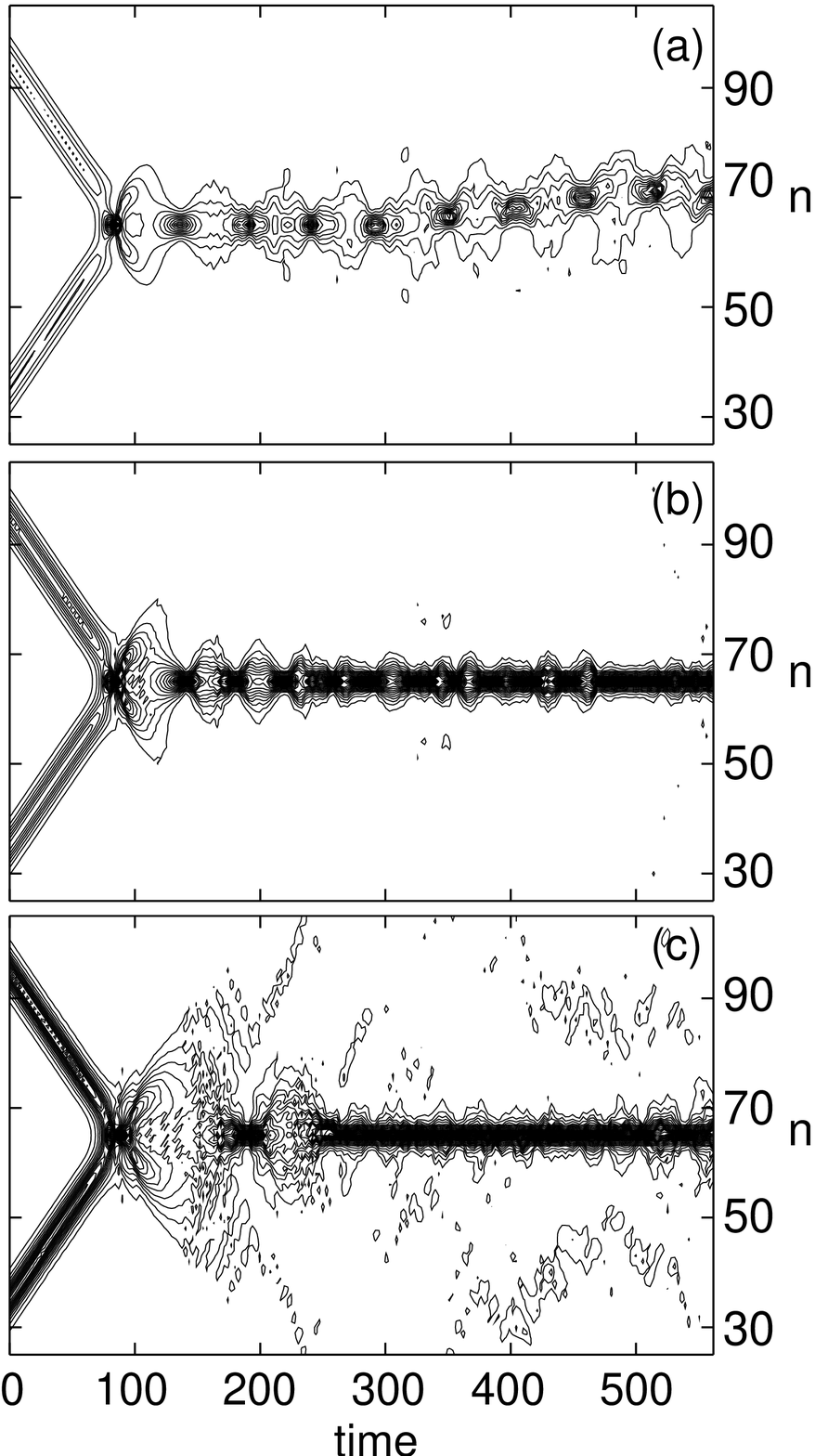} &
\end{tabular} \end{center}
\caption{Contour plots showing the evolution of the lattice field
$|\Phi _{n}|$ in three cases of collisions between identical
solitons moving in opposite directions. The soliton's
frequency is $\protect\omega _{b}=-2.11$, and $|\protect\mu |=0.6$
(a), $|\protect\mu |=0.8$ (b), and $|\protect\mu  |=0.9$ (c).}
\label{fig:8}
\end{figure}

The most notable and generic feature of the collision manifests itself 
in the merger scenario. When the cores of the mobile solitons collide, 
sudden delocalization is first observed, with transfer of energy from 
the collision point to adjacent lattice sites. Then, almost all the 
energy is collected back at the collision spot, and thus a single 
localized state emerges. An example of the collision is shown in Fig. 
\ref{fig:8}. This scenario was observed in all simulations of the 
collisions.
The appearance of pulsons as the product of soliton collisions, as 
well as the fact that they also appear as asymptotic states of the 
evolution of perturbed unstable solitons (see Section \ref{subsec:Stability}), 
shows the ubiquity of this type of localized excitations in the 
present model. 

\section{Conclusion}
\label{sec:Conclusion}

We have introduced a combination of the DNLS and Ablowitz-Ladik models with
competing nonlinearities (self-focusing onsite and self-defocusing
intersite cubic terms). The model may describe a condensate of bosonic
dipoles trapped in a strong lattice potential. First, it was shown that the
continuum counterpart of the model gives rise to solitons in a finite band
of frequencies, with broad, small-amplitude NLS-like solitons at one edge
of the band, and an exact solution in the form of a peakon at the other.
Numerical analysis of the discrete system yields a similar family of
solitons, which also includes a peakon; however, the family continues 
beyond the peakon, in the form of discrete cuspons. Stability of the 
lattice solitons was investigated by means of the computation of 
their Floquet spectra, and in direct simulations. 
It was found that only a small part of the soliton family is unstable; 
the evolution of the unstable solitons leads asymptotically to pulsons, 
{\em i.e.} localized solutions where the width oscillates.\ 
In particular, peakons and cuspons are stable. Additionally, 
it was found that the Vakhitov-Kolokolov criterion precisely 
explains the stability and instability of regular solitons, but
does not apply to cuspons, for which it erroneously predicts instability.
Bound states of identical solitons were also investigated, revealing a 
stability exchange: the in-phase and out-of-phase bound states, which are
unstable and stable, respectively, in the DNLS limit, exchange their 
stability character exactly at the point where the bound solitons are 
peakons.

We have numerically computed mobile solitons with the same high precision as 
for the static ones. Their structure is that of an exponentially localized 
core on top of an extended background (which is a superposition of a finite 
number of extended plane waves). The amplitude of the background is correlated 
with the height of the computed Peierls-Nabarro barrier. The mobile solitons 
exist up to a critical value of the strength of the self-defocusing intersite 
nonlinearity, which is lower than (but close to) the corresponding value for 
which the peakon state exist. Collisions between solitons always lead to their 
merger into a pulson state.

\section*{Acknowledgements}
The authors are pleased to acknowledge Franz G. Mertens for useful 
discussions. J.G.-G. appreciates hospitality of the Center 
for Nonlinear Studies at Los Alamos National Laboratory and 
acknowledges financial support from the MECyD through a FPU grant.
B.A.M. appreciates hospitality of the Department of Condensed Matter 
Physics at the University of Zaragoza. Work at Los Alamos is supported 
by the USDoE. This work was partially supported by MCyT (Project 
No. BFM2002 00113), DGA and BIFI.


\begin{thebibliography}{99}
\bibitem{Footnote} Discrete solitons are also widely referred to as discrete 
breathers because of the ocillatory dynamics of the real and imaginary part of 
the complex variables $\{\Phi _{n}\}$.
\end{thebibliography}
\end{document}